\documentclass{article}[12pt]
\usepackage{amsmath,amssymb}
\usepackage{graphicx}%
\usepackage{verbatim}

\newcommand{\Frac}[2]{{\displaystyle \frac{#1}{#2}}}

\begin{document}

\begin{center}
{\bf \Large Qualitative and Numerical Analysis of \\[6pt]
a Cosmological Model Based on an Asymmetric\\[6pt]
Scalar Doublet with Minimal Couplings.\\[6pt]
I. Qualitative Analysis of the Model} \\[12pt]
Yu. G. Ignat'ev and I. A. Kokh\\
N. I. Lobachevsky Institute of Mathematics and Mechanics of Kazan Federal University,\\
Kremleovskaya str., 35, Kazan, 420008, Russia.
\end{center}


\begin{abstract}

A qualitative analysis of a cosmological model based on the asymmetric scalar doublet classical + phantom
scalar field with minimal interaction is performed. It is shown that depending on the parameters of the model,
the corresponding dynamical system can have 1, 3, or 9 stationary points corresponding to attractive or
repulsive centers (1--5) and saddle points (0--4). A physical analysis of the model is performed.\\


\end{abstract}

\section{The basic equations of a cosmological model based on an asymmetric scalar doublet}
\subsection{Lagrange function and interaction potential}
In [1] a cosmological model, based on an asymmetric scalar doublet, that is, a system consisting of two scalar
fields -- a classical field, $\Phi$, and a phantom field, $\varphi$ -- was proposed and partially investigated. A qualitative analysis was
also performed for the case of free scalar fields interacting with each other only via gravitation, and it was conjectured
that even a weak phantom scalar field might be able to have a substantial effect on the dynamics of the cosmological
model. However, in [1], first of all, some errors were made in the course of the qualitative analysis of the dynamical
system, and on top of that, no systematic numerical modeling was performed to elucidate unique features of the
cosmological model. We attempt here to correct the indicated shortcomings, and we will also give an energy-based
interpretation of the obtained results.
The Lagrange function of a scalar doublet consisting of a classical and a phantom scalar field with self-action
in Higgs form with minimal coupling has the form [1]

\begin{equation} \label{1}
L=\frac{1}{8\pi } (g^{ik} \Phi _{,i} \Phi _{,k} -2V(\Phi ))-\frac{1}{8\pi } (g^{ik} \varphi _{,i} \varphi _{,k} +2v(\varphi )), \end{equation}
where

\[V(\Phi )=-\frac{\alpha }{4} \left(\Phi ^{2} -e\frac{m^{2} }{\alpha } \right)^{2} ;{\rm \; \; \; }v(\varphi )=-\frac{\beta }{4} \left(\varphi ^{2} -\varepsilon \frac{{\rm m}^{2} }{\beta } \right)^{2} \]
is the Higgs potential energy of the corresponding scalar fields, $\alpha$ and $\beta$  are their self-action constants  $m$ and $\rm m$  are
the masses of the quanta. Introducing the summed potential
\begin{equation} \label{2}
 U(\Phi ,\varphi )=V(\Phi )+v(\varphi )=-\frac{\alpha }{4} \left(\Phi ^{2} -e\frac{m^{2} }{\alpha } \right)^{2} -\frac{\beta }{4} \left(\varphi ^{2} -\varepsilon \frac{{\rm m}^{2} }{\beta } \right)^{2} \equiv U(e,\varepsilon ,\alpha ,\beta ;\Phi ,\varphi ),
\end{equation}
it is possible to draw the following conclusions:

1. The potential $U(\Phi ,\varphi )$ possesses the following symmetries:
\begin{equation}\label{3}
U(\pm \Phi ,\pm \varphi )=U(\Phi ,\varphi );
\end{equation}
\begin{equation}\label{4}
U(-e,-\varepsilon ,-\alpha ,-\beta ;\Phi ,\varphi )=-U(e,\varepsilon ,\alpha ,\beta ;\Phi ,\varphi ).
\end{equation}

2. For $\{ e=1,\varepsilon =1\} $ the function $U(\Phi ,\varphi )$ has an absolute maximum at the origin of the phase plane $\{ \Phi ,\varphi \} $ $M_{0} (0,0)$, and for $\{ e=-1,\varepsilon =-1\} $ it has an absolute minimum, i.e., everywhere that $e\varepsilon =1$, and for $e\varepsilon =-1$ it has a conditional extremum (saddle point).

3. For $e\alpha >0$ and $\varepsilon \alpha <0$ the function $U(\Phi ,\varphi )$ has an absolute maximum at the points $M_{10} (-m/\sqrt{e\alpha } ,0)$ and $M_{20} (m/\sqrt{e\alpha } ,0)$ for  $\alpha <0$ (i.e., $\alpha <0,e=-1,\varepsilon =+1$) and an absolute minimum at these points for  $\alpha >0$ (i.e., $\alpha >0,e=+1,\varepsilon =-1$), i.e., everywhere that $e\varepsilon =-1$; for $e\alpha >0$ and $\varepsilon \alpha >0$ it has a conditional
extremum at these points (saddle points), i.e., everywhere that $e\varepsilon =1$.

4. For $\varepsilon \beta >0$ and $e\beta <0$ function $U(\Phi ,\varphi )$ has an absolute maximum at the points $M_{01} (0,-m/\sqrt{\varepsilon \beta } )$ and $M_{02} (0,m/\sqrt{\varepsilon \beta }) $ for  $\beta <0$ (i.e., $\beta <0,e=+1,\varepsilon =-1$) and an absolute minimum at these points for  $\beta >0$ (i.e.,$\beta >0,e=-1,\varepsilon =+1$), i.e., everywhere that $e\varepsilon =-1$; for $\varepsilon \beta >0$ and $e\beta >0$ it has a conditional extremum at these points (saddle points), i.e., for $e\varepsilon =1$.

5. For $e\alpha >0$, $\varepsilon \beta >0$ and $\alpha \beta >0$ the function $U(\Phi ,\varphi )$ has an absolute maximum at the points $M_{11} (-m/\sqrt{e\alpha } ,-m/\sqrt{\varepsilon \beta } )$, $M_{12} (-m/\sqrt{e\alpha } ,m/\sqrt{\varepsilon \beta } )$,\linebreak $M_{21} (m/\sqrt{e\alpha } ,-m/\sqrt{\varepsilon \beta } )$ and $M_{22} (m/\sqrt{e\alpha } ,m/\sqrt{\varepsilon \beta } )$ for  $\alpha >0$ (i.e., $\alpha >0,\beta >0,e=1,\varepsilon =1$) and an absolute minimum at these points for  $\alpha <0$ (i.e., $\alpha <0,\beta <0,e=-1,\varepsilon =-1$), at all of these points $e\varepsilon =1$; for $\alpha \beta <0$ it has a conditional extremum at these
points (saddle points), i.e., for $e\varepsilon =-1$.

Typical graphs of the potential function $U(\Phi ,\varphi )$, corresponding to the two opposite cases described in paragraph 5 are shown in Figs. 1 and 2.

\begin{figure}[h!]
 \centerline{\includegraphics[width=8cm]{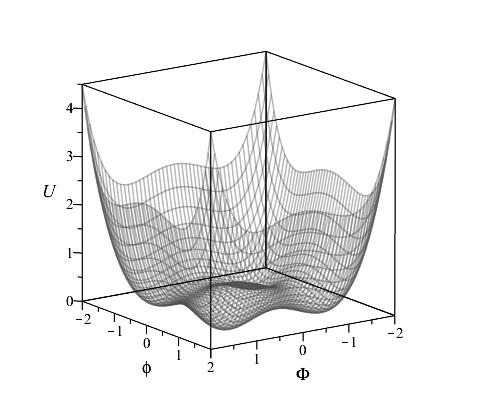}\label{Fig1}} \caption{Graph of the potential  $U(-1,-1,-1,-1;\Phi ,\varphi )$ .
 }
\end{figure}
 \begin{figure}[h!]
 \centerline{\includegraphics[width=8cm]{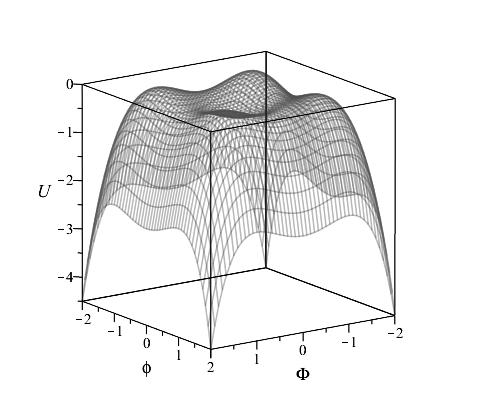}\label{Fig2}} \caption{Graph of the potential $U(1,1,1,1;\Phi ,\varphi )$ .
 }
\end{figure}

 In Fig. 1 it is possible to discern four minima and one central maximum, and
also four saddle points. In Fig. 2 it is possible to discern four maximum, one central minimum, and four saddle points.
The latter figure, naturally, is obtained by a mirror reflection of the first about the plane $U=0$.

Thus, taking into account the fact that the stationary points of the dynamical system with Lagrange function of the form given by Eq. \eqref{1} coincides with the stationary points of the potential $U(\Phi ,\varphi )$, we can safely assert the following. Depending on the signs of the parameters $\{ e,\varepsilon ,\alpha ,\beta \} $ of the potential \linebreak $U(e,\varepsilon ,\alpha ,\beta ;\Phi ,\varphi )$, the corresponding dynamical system can have 1, 3, or 9 stationary points, among which there are attractive points (absolute minimum), repulsive points (absolute maximum) and saddle points (conditional extremum). This result is in complete agreement with the conclusions of [1], in which it was obtained with the help of the qualitative theory of dynamical systems.

\subsection{Equations of the cosmological model}
The energy-momentum tensor of the scalar field relative to the Lagrange function (Eq. \eqref{1}) takes the standard form:

\begin{equation} \label{5} T_{ik} =\frac{1}{8\pi } (2\Phi _{,i} \Phi _{,k} -g_{ik} \Phi _{,j} \Phi ^{,j} +2V(\Phi )g_{ik} )-\frac{1}{8\pi } (2\varphi _{,i} \varphi _{,k} +g_{ik} \varphi _{,j} \varphi ^{,j} -2v(\varphi )g_{ik} ).
\end{equation}
Variation of the Lagrange function (Eq. \eqref{1}) leads to the following field equations:

\[\begin{array}{l} {\square \Phi +V'(\Phi )=0;{\rm \; }} \\ {\square \varphi +v'(\varphi )=0.} \end{array}\]
Renormalizing the Lagrange function (Eq. \eqref{1}) and adding a constant to it [2], we reduce it to the form

\begin{equation} \label{6} L=\frac{{\rm 1}}{8\pi } \left(g^{ik} \Phi _{,i} \Phi _{,k} -em^{2} \Phi ^{2} +\frac{\alpha }{2} \Phi ^{4} \right)-\frac{{\rm 1}}{8\pi } \left(g^{ik} \varphi _{,i} \varphi _{,k} +\varepsilon {\rm m}^{2} \varphi ^{2} -\frac{\beta }{2} \varphi ^{4} \right), \end{equation}
the corresponding renormalization of the energy-momentum tensor gives

\begin{equation} \label{7} \begin{array}{l} {T_{ik} =\frac{1}{8\pi } \left(2\Phi _{,i} \Phi _{,k} -g_{ik} \Phi _{,j} \Phi ^{,j} +g_{ik} em^{2} \Phi ^{2} -g_{ik} \frac{\alpha }{2} \Phi ^{4} \right)+} \\[12pt]  {-\frac{{\rm 1}}{8\pi } \left(2\varphi _{,i} \varphi _{,k} -g_{ik} \varphi _{,j} \varphi ^{,j} -g_{ik} \varepsilon {\rm m}^{2} \varphi ^{2} +g_{ik} \frac{\beta }{2} \varphi ^{4} \right).} \end{array} \end{equation}

Applying the standard variational procedure to the Lagrange function in the form given by Eq. \eqref{6}, we obtain the equations of the free classical and phantom fields:

\begin{equation} \label{8} \square \Phi +m_{*}^{2} \Phi =0;{\rm \; } \end{equation}
\begin{equation} \label{9} \square \varphi +{\rm m}_{*}^{2} \varphi =0. \end{equation}
where $m_{*}^{} $ and ${\rm m}_{*}^{} $ are the effective masses of the scalar bosons:

\begin{equation} \label{10} \begin{array}{l} {m_{*} ^{2} =em^{2} -\alpha \Phi ^{2} ;} \\[12pt]  {{\rm m}_{*} ^{2} =-\varepsilon {\rm m}^{2} +\beta \varphi ^{2} .} \end{array} \end{equation}

 The Einstein equations with the cosmological term\footnote{ The Planck system of units is used:  $G=c=h=1$;  the Ricci tensor is obtained by contraction of the first and fourth indices  $R_{ik} =R_{ikj}^{j} $ ; the metric has the signature (-1, -1, -1, +1).} have the form

\begin{equation} \label{11} R^{ik} -\frac{1}{2} Rg^{ik} =\lambda g^{ik} +8\pi T^{ik} , \end{equation}
where $\lambda\geq0$ is the cosmological constant. Next, let us consider the self-consistent system of equations \eqref{8}, \eqref{9}, \eqref{11}, based on a free asymmetric scalar doublet, together with the spatially-flat Friedmann metric:

\begin{equation} \label{12} ds^{2} =dt^{2} -a^{2} (t)(dx^{2} +dy^{2} +dz^{2} ), \end{equation}
where we set  $\Phi =\Phi (t)$ and $\varphi =\varphi (t).$

Here the energy-momentum tensor (Eq. \eqref{7}) takes on the structure of the
energy-momentum tensor of an isotropic fluid with energy density $\mathcal{E} $ and pressure~$p$:

\begin{equation} \label{13} \mathcal{E}(t)=\mathcal{E}_{c} +\mathcal{E}_{f} ;{\rm \; \; }p=p_{c} +p_{f} ; \end{equation}

\begin{eqnarray} \label{14} \nonumber
\mathcal{E}_{c} =\frac{1}{8\pi } \left(\dot{\Phi }^{2} +em^{2} \Phi ^{2} -\frac{\alpha }{2} \Phi ^{4} \right);&\mathcal{E}_{f} =\frac{1}{8\pi } \left(-\dot{\varphi }^{2} +\varepsilon {\rm m}^{2} \varphi ^{2} -\frac{\beta }{2} \varphi ^{4} \right);\\
 p_{c} =\frac{1}{8\pi } \left(\dot{\Phi }^{2} -em^{2} \Phi ^{2} +\frac{\alpha }{2} \Phi ^{4} \right);&{\rm }p_{f} =\frac{1}{8\pi } \left(-\dot{\varphi }^{2} -\varepsilon {\rm m}^{2} \varphi ^{2} +\frac{\beta }{2} \varphi ^{4} \right), \end{eqnarray}
where $\dot{f}{\rm \equiv }df/dt$. Here the identity

\begin{equation} \label{15} \mathcal{E}+p\equiv \frac{1}{4\pi } \dot{\Phi }^{2} -\frac{1}{4\pi } \dot{\varphi }^{2}
\end{equation}
is satisfied. The investigated system consists of one Einstein equation

\begin{equation} \label{16} 3\frac{\dot{a}^{2} }{a^{2} } \equiv 3H^{2} =\left(\dot{\Phi }^{2} +em^{2} \Phi ^{2} -\frac{\alpha }{2} \Phi ^{4} \right)-\left(\dot{\varphi }^{2} -\varepsilon {\rm m}^{2} \varphi ^{2} +\frac{\beta }{2}  ^{4} \right)+\lambda  \end{equation}
and two scalar-field equations:

\begin{equation} \label{17} \ddot{\Phi }+3\frac{a}{a} \dot{\Phi }+m_{*}^{2} \Phi =0, \end{equation}

\begin{equation} \label{18} \ddot{\varphi }+3\frac{\dot{a}}{a} \dot{\varphi }+{\rm m}_{*}^{2} \varphi =0. \end{equation}

Substituting the expressions for the effective masses $m_{*}^{} $ and ${\rm m}_{*}^{}$ given by Eqs. \eqref{10} into Eqs. \eqref{14} and \eqref{15},
we obtain the final form of our system of equations:

\begin{equation} \label{19} 3\frac{\dot{a}^{2} }{a^{2} } =\left(\dot{\Phi }^{2} +em^{2} \Phi ^{2} -\frac{\alpha }{2} \Phi ^{4} \right)-\left(\dot{\varphi }^{2} -\varepsilon {\rm m}^{2} \varphi ^{2} +\frac{\beta }{2} \varphi ^{4} \right)+\lambda ; \end{equation}

\begin{equation} \label{20} \ddot{\Phi }+3\frac{\dot{a}}{a} \dot{\Phi }+em_{}^{2} \Phi -\alpha \Phi ^{3} =0; \end{equation}

\begin{equation} \label{21} \ddot{\varphi}+3\frac{\dot{a}}{a} \dot{\varphi }-\varepsilon {\rm m}_{}^{2} \varphi +\beta \varphi ^{3} =0. \end{equation}

The Hubble constant and the invariant cosmological acceleration have the form

\[H(t)=\frac{\dot{a}}{a} \ge 0;{\rm }\Omega (t)=\frac{a\ddot{a}}{\dot{a}^{2} } \equiv 1+\frac{\dot{H}}{H^{2} } ,\]
where the cosmological acceleration, which is an invariant, is expressed with the help of the coefficient of barotropy $\chi =p/\mathcal{E}$:

\begin{equation} \label{22} \Omega =-\frac{1}{2} (1+3\chi ). \end{equation}

\section{Qualitative analysis of the cosmological model}
\subsection{Reduction of the system of equations to normal form}

Turning now to the dimensionless Compton time $mt=\tau$ ($m\ne 0$)  and carrying out the standard substitution of variables

\begin{equation} \label{23} \Phi '=Z(\tau ),{\rm }\varphi '=z(\tau ),{\rm }(f'\equiv df{\rm /d}\tau ), \end{equation}
we reduce the Einstein equation (Eq. \eqref{19}) to dimensionless form:

\begin{equation} \label{24} H'^{2}_{m}=\frac{1}{3} \left[\left(Z^{2} +e\Phi ^{2} -\frac{\alpha _{m} }{2} \Phi ^{4} \right)-\left(z^{2} -\varepsilon \mu ^{2} \varphi ^{2} +\frac{\beta _{m} }{2} \varphi ^{4} \right)+\lambda _{m} \right],\end{equation}
and the field equations (Eqs. \eqref{20} and \eqref{21}) to the form of a normal autonomous system of ordinary differential equations in the four-dimensional phase space ${\mathbb R}_{4} :\{ \Phi ,Z,\varphi ,z\} $

\begin{equation} \label{25} \begin{array}{l} {\Phi '=Z;} \\ {Z'=-\sqrt{3} Z\sqrt{\left(Z^{2} +e\Phi ^{2} -\Frac{\alpha _{m} }{2} \Phi ^{4} \right)-\left(z^{2} -\varepsilon \mu ^{2} \varphi ^{2} +\Frac{\beta _{m} }{2} \varphi ^{4} \right)+\lambda _{m} } -e\Phi +\alpha _{m} \Phi ^{3} ;} \\ {\varphi '=z;} \\ {z'=-\sqrt{3} z\sqrt{\left(Z^{2} +e\Phi ^{2} -\Frac{\alpha _{m} }{2} \Phi ^{4} \right)-\left(z^{2} -\varepsilon \mu ^{2} \varphi ^{2} +\Frac{\beta _{m} }{2} \varphi ^{4} \right)+\lambda _{m} } +\varepsilon \mu ^{2} \varphi -\beta _{m} \varphi ^{3} ,} \end{array} \end{equation}
Here we have introduced the following notation:

\[\lambda _{m} \equiv \frac{\lambda }{m^{2} } ;{\rm }\alpha _{m} \equiv \frac{\alpha }{m^{2} } ;{\rm }\beta _{m} \equiv \frac{\beta }{m^{2} } ;{\rm }\mu \equiv \frac{{\rm m}}{m} .\]
Here

\[\frac{a}{a} \equiv \Lambda '=H_{m} \equiv \frac{H}{m} ;{\rm }\Omega =\frac{aa''}{a'^{2} } \equiv 1+\frac{h}{h^{2} } ,\]
where

\[\Lambda =\ln a(\tau ).\]

In this notation, all of the quantities of the problem are dimensionless; the time $\tau $ is measured in Compton units
referenced to the classical scalar field. To start with, we note that the conclusions of the qualitative theory regarding the
stationary points of the dynamical system and their character cannot differ from conclusions based on an analysis of the
potential function. However, the qualitative theory provides a more detailed description of the behavior of the
dynamical system near the stationary points.
In order for system of differential equations \eqref{25} to have a real solution, it is necessary that the expression
inside the radical in the equations be nonnegative, i.e., that the effective energy of the system, taking the cosmological
constant into account, be nonnegative:
\begin{equation}\label{26}
\left(Z^{2} +e\Phi ^{2} -\frac{\alpha _{m} }{2} \Phi ^{4} \right)+\left(-z^{2} +\varepsilon \mu ^{2} \varphi ^{2} -\frac{\beta _{m} }{2} \varphi ^{4} \right)+\lambda _{m} \ge 0.
\end{equation}
Inequality \eqref{26} can lead to violation of simple connection of the phase space and formation in it of closed lacunas,bounded by surfaces with zero effective energy\footnote{We will address the question of the motion of the system near these lacunas in a subsequent paper.}. To reduce system \eqref{25} to the standard notation of the qualitative
theory (for example, see [2])

\[\frac{dx_{i} }{d\tau } =F_{i} (x_{1} ,\ldots ,x_{n} ),{\rm }i=\overline{1,n}\]
we adopt the following notation:

\begin{equation} \label{27} \begin{array}{l} {\Phi =x;\quad \varphi =y;\quad F_{1} \equiv P=Z;\quad F_{3} \equiv p=z;} \\[12pt] {F_{2} \equiv Q=-\sqrt{3} Z\sqrt{\left(Z^{2} +ex^{2} -\Frac{\alpha _{m} }{2} x^{4} \right)-\left(z^{2} -\varepsilon \mu ^{2} y^{2} +\Frac{\beta _{m} }{2} y^{4} \right)+\lambda _{m} } -ex+\alpha _{m} x^{3} ;} \\[12pt]  {F_{4} \equiv q=-\sqrt{3} z\sqrt{\left(Z^{2} +ex^{2} -\Frac{\alpha _{m} }{2} x^{4} \right)-\left(z^{2} -\varepsilon \mu ^{2} y^{2} +\Frac{\beta _{m} }{2} y^{4} \right)+\lambda _{m} } +\varepsilon \mu ^{2} y-\beta _{m} y^{3} .} \end{array} \end{equation}
The corresponding normal system of equations in the standard notation has the form

\begin{equation} \label{28} x'=P;\quad Z'=Q;\quad y'=p;\quad z'=q. \end{equation}
The necessary condition for the real solution (inequality \eqref{26}) is rewritten in the form

\begin{equation} \label{29} \left(Z^{2} +ex^{2} -\frac{\alpha _{m} }{2} x^{4} \right)-\left(z^{2} -\varepsilon \mu ^{2} y^{2} +\frac{\beta _{m} }{2} y^{4} \right)+\lambda _{m} \ge 0. \end{equation}

\subsection{Singular points of the dynamical system}

The singular points of the dynamical system are determined by a system of algebraic equations (for example, see [2, 3]):

\begin{equation} \label{30} M:\quad F_{i} (x_{1} ,\ldots ,x_{n} )=0,\quad i=\overline{1,n}. \end{equation}
According to Eqs. \eqref{27} and \eqref{30}, these points are determined by the system of equations

\begin{equation} \label{31} \begin{array}{l} {Z=0,\quad z=0;} \\[12pt] {x(e-\alpha _{m} x^{2} )=0;} \\[12pt] {y(\varepsilon \mu ^{2} -\beta _{m} y^{2} )=0.} \end{array} \end{equation}

Thus, as we indicated above, dynamical system \eqref{27} has nine singular points.

\begin{enumerate}
\item  $M_0$: For arbitrary values of $\alpha _{m} $  and $\beta _{m} $   system of algebraic equations \eqref{30}  always has the trivial solution
\begin{equation} \label{32} x=0;\quad Z=0;\quad y=0;\quad z=0\Rightarrow M_{0} :(0,0,0,0). \end{equation}
Substituting the obtained solution (Eq. \eqref{32}) into condition \eqref{29}, we obtain a necessary condition for the real solutions at
the singular point:
\begin{equation} \label{33} \lambda _{m} \ge 0. \end{equation}
\item  $M_{01}, M_{02}$ : For arbitrary values of $\alpha _{m} $  and $\varepsilon \beta _{m} >0$  we have two more solutions, which are symmetric
in $\varphi $ :
\begin{equation} \label{34} \begin{array}{l} {x=0;\quad Z=0;\quad y_{\pm } =\pm \frac{\mu }{\sqrt{\varepsilon \beta _{m} } } ;\quad z=0\quad \Rightarrow} \\[12pt]
{M_{01} (0,0,|y_{\pm } |,0);\quad M_{02} (0,0,-|y_{\pm } |,0).} \end{array} \end{equation}
A necessary condition for the real solutions at the singular points $M_{01}, M_{02}$ is:
\begin{equation} \label{35} \frac{\mu ^{4} }{2\beta _{m} } +\lambda _{m} \ge 0. \end{equation}
\item  $M_{10}, M_{20}$ : For arbitrary $\beta _{m} $  and  $e\alpha _{m} >0$  we have two more solutions, which are symmetric in $\Phi $ :

\begin{equation} \label{36} \begin{array}{l} { x_{\pm } =\pm \frac{1}{\sqrt{e\alpha _{m} } } ;\quad Z=0;\quad y=0;\quad z=0\quad \Rightarrow} \\[12pt]
{M_{10} (|x_{\pm } |,0,0,0);\quad M_{20} (-|x_{\pm } |,0,0,0).} \end{array} \end{equation}

A necessary condition for the real solutions at the singular points $M_{10}, M_{20}$ is:

\begin{equation} \label{37} \frac{1}{2\alpha _{m} } +\lambda _{m} \ge 0. \end{equation}

\item  $M_{12}, M_{21}, M_{11}, M_{22}$ : For  $e\alpha _{m} >0$  and $\varepsilon \beta _{m} >0$  we have four more solutions, which are symmetric in $\Phi $  and  $\varphi $ :
\begin{equation} \label{38} \begin{array}{l} {x_{\pm } =\pm \Frac{1}{\sqrt{e\alpha _{m} } } ;\quad Z=0;\quad y_{\pm } =\pm \Frac{\mu }{\sqrt{\varepsilon \beta _{m} } } ;\quad z=0\quad \Rightarrow } \\[12pt] {M_{11} (|x_{\pm } |,0,|y_{\pm } |,0);{\rm }M_{12} (|x_{\pm } |,0,-|y_{\pm } |,0);} \\[12pt] {M_{21} (-|x_{\pm } |,0,|y_{\pm } |,0);M_{22} (-|x_{\pm } |,0,-|y_{\pm } |,0).} \end{array} \end{equation}

A necessary condition for the real solutions at the singular points $M_{12}, M_{21}, M_{11}, M_{22}$ is:
\begin{equation} \label{39} \frac{1}{2\alpha _{m} } +\frac{\mu ^{4} }{2\beta _{m} } +\lambda _{m} \ge 0. \end{equation}
Let us investigate the character of the obtained singular points. The matrix of dynamical system \eqref{27} for $Z=z=0$ has the form:

\begin{equation} \label{40} A=\left\| \displaystyle\frac{\partial F_{i} }{\partial x_{k} } \right\| =\left(\begin{array}{cccc} {0} & {1} & {0} & {0} \\ \displaystyle{\frac{\partial Q}{\partial x} } &\displaystyle {\frac{\partial Q}{\partial Z} } & {0} & {0} \\ {0} & {0} & {0} & {1} \\ {0} & {0} & \displaystyle{\frac{\partial q}{\partial y} } & \displaystyle{\frac{\partial q}{\partial z} } \end{array}\right). \end{equation}

The determinant of this block-diagonal matrix is equal to

\begin{equation} \label{41} \Delta (A)=\frac{\partial Q}{\partial x} \frac{\partial q}{\partial y} . \end{equation}
\end{enumerate}
\subsection{Characteristic equation and qualitative analysis of the zero singular point $M_{0}$}

In order for the dynamical system to allow phase trajectories to arrive at singular points, it is necessary that the coordinates of these points have real values, namely, that the condition $\lambda _{m} \ge 0$ be fulfilled. In this case, the matrix of system \eqref{27} at the zero singular point with coordinates \eqref{32} for arbitrary $\alpha _{m} $ and $\beta _{m} $ takes the following form:

\begin{equation} \label{42} A_{0} \equiv A(M_{0} )=\left. \left(\begin{array}{cccc} {0} & {1} & {0} & {0} \\ \displaystyle{\frac{\partial Q}{\partial x} } & \displaystyle{\frac{\partial Q}{\partial Z} } & {0} & {0} \\ {0} & {0} & {0} & {1} \\ {0} & {0} & \displaystyle{\frac{\partial q}{\partial y} } & \displaystyle{\frac{\partial q}{\partial z} } \end{array}\right){\rm }\right|_{M_{0} } =\left(\begin{array}{cccc} {0} & {1} & {0} & {0} \\ {-e} & {-\sqrt{3\lambda _{m} } } & {0} & {0} \\ {0} & {0} & {0} & {1} \\ {0} & {0} & {\varepsilon \mu ^{2} } & {-\sqrt{3\lambda _{m} } } \end{array}\right), \end{equation}
and its determinant

\begin{equation} \label{43} \Delta (A_{0} )=-e\varepsilon \mu ^{2} . \end{equation}
The characteristic equation for the matrix $A_{0} $ has the form:

\[(k^{2} +k\sqrt{3\lambda _{m} } +e)(k^{2} +k\sqrt{3\lambda _{m} } -\varepsilon \mu ^{2} )=0,\]
thus, the eigenvalues of the matrix are equal to

\begin{equation} \label{44} \begin{array}{l}
{k_{1} (M_{0} )=-\frac{1}{2} \sqrt{3\lambda _{m} } +\frac{1}{2} \sqrt{3\lambda _{m} +4\varepsilon \mu ^{2} };}\quad {k_{2} (M_{0} )=-\frac{1}{2} \sqrt{3\lambda _{m} } -\frac{1}{2} \sqrt{3\lambda _{m} +4e\varepsilon \mu ^{2} }}; \\[12pt]
{k_{3} (M_{0} )=-\frac{1}{2} \sqrt{3\lambda _{m} } +\frac{1}{2} \sqrt{3\lambda _{m} -4e} ;}\quad\quad{k_{4} (M_{0} )=-\frac{1}{2} \sqrt{3\lambda _{m} } -\frac{1}{2} \sqrt{3\lambda _{m} -4e}},\end{array}
\end{equation}

here

\begin{equation} \label{45} \begin{array}{l} {k_{1} (M_{0} )\cdot k_{2} (M_{0} )=-\varepsilon \mu ^{2} ;{\rm \; \; }k_{3} (M_{0} )\cdot k_{4} (M_{0} )=e;{\rm \; \; }} \\[12pt]
 {k_{1} (M_{0} )\cdot k_{2} (M_{0} )\cdot k_{3} (M_{0} )\cdot k_{4} (M_{0} )=\Delta (A_{0} ).} \end{array} \end{equation}

Therefore:

\begin{enumerate}
\item  for $e=+1$ $k_{3} $ and $k_{4} $ are either complex conjugate numbers or real numbers with identical signs; for $e=-1$  $k_{3} $ and $k_{4} $ are real numbers with different signs;

\item  for $\varepsilon =+1$  $k_{1} $ and $k_{2} $ are real numbers with different signs; for $\varepsilon =-1$ $k_{1} $ and $k_{2} $ are either complex conjugate numbers or real numbers with identical signs.
\end{enumerate}

\subsection{Characteristic equation and qualitative analysis near the singular points $M_{01} ,M_{02} $}

In order for the dynamical system to allow phase trajectories to arrive at singular points, it is necessary that the
coordinates of these points have real values, i.e., that

\[\frac{\mu ^{4} }{2\beta _{m} } +\lambda _{m} \ge 0.\]

In this case the matrix of system \eqref{27} at the singular points defined by Eq. \eqref{34} for  $\varepsilon \beta _{m} >0$ has the form

\begin{equation} \label{46} A_{01} \equiv A(M_{01} )=A(M_{02} )={\rm }\left(\begin{array}{cccc} {0} & {1} & {0} & {0} \\ {-e} & {-\sqrt{3\left(\displaystyle\frac{\mu ^{4} }{2\beta _{m} } +\lambda _{m} \right)} } & {0} & {0} \\ {0} & {0} & {0} & {1} \\ {0} & {0} & {-2\varepsilon \mu ^{2} } & {-\sqrt{3\left(\displaystyle\frac{\mu ^{4} }{2\beta _{m} } +\lambda _{m} \right)} } \end{array}\right){\rm \; }, \end{equation}

and its determinant

\begin{equation} \label{47} \Delta (A_{01} )=2e\varepsilon \mu ^{2} . \end{equation}

In this case, the characteristic equation for the matrix $A_{01}$ takes the following form

\[\left(k^{2} +k\sqrt{3\left(\frac{\mu ^{4} }{2\beta _{m} } +\lambda _{m} \right)} +2\varepsilon \mu ^{2} \right)\left(k^{2} +k\sqrt{3\left(\frac{\mu ^{4} }{2\beta _{m} } +\lambda _{m} \right)} +e\right)=0.\]

Thus, we have found the eigenvalues of the matrix:
\begin{equation} \label{48} \begin{array}{l} {k_{1} (M_{01} )=-\frac{1}{2} \sqrt{3u} +\frac{1}{2} \sqrt{3u-8\varepsilon \mu ^{2} } ;\quad k_{2} (M_{01} )=-\frac{1}{2} \sqrt{3u} -\frac{1}{2} \sqrt{3u-8\varepsilon \mu ^{2} } ;} \\[12pt] {k_{3} (M_{01} )=-\frac{1}{2} \sqrt{3u} +\frac{1}{2} \sqrt{3u-4e} ;\quad\quad k_{4} (M_{01} )=-\frac{1}{2} \sqrt{3u} -\frac{1}{2} \sqrt{3u-4e} ,} \end{array} \end{equation}

where  $u=\displaystyle\frac{\mu ^{4} }{2\beta _{m} } +\lambda _{m}.$

Here

\begin{equation} \label{49_} \begin{array}{l} {k_{1} (M_{01} )\cdot k_{2} (M_{01} )=2\varepsilon \mu ^{2} ;{\rm \; \; }k_{3} (M_{01} )\cdot k_{4} (M_{01} )=e;{\rm \; \; }} \\[12pt]  {k_{1} (M_{01} )\cdot k_{2} (M_{01} )\cdot k_{3} (M_{01} )\cdot k_{4} (M_{01} )=\Delta (A_{01} ).} \end{array} \end{equation}

Therefore:

\begin{enumerate}
\item  for $e=+1$  $k_{3} $ and $k_{4} $ are either complex conjugate numbers or real numbers with identical signs; for $e=-1$  $k_{3} $ and $k_{4} $ are real numbers with different signs;

\item  for $\varepsilon =+1$ $k_{1} $ and $k_{2} $ are either complex conjugate numbers or real numbers with identical signs; for $\varepsilon =-1$ $k_{1} $ and $k_{2}$ are real numbers with different signs.
\end{enumerate}

\subsection{Characteristic equation and qualitative analysis near the singular points $M_{10} ,M_{20} $ }

In order for the dynamical system to allow phase trajectories to arrive at singular points, it is necessary that the
coordinates of these points have real values, i.e., that

\[\frac{1}{2\alpha _{m} } +\lambda _{m} \ge 0,\]

In this case, the matrix of system \eqref{27} at the singular points \eqref{36} for  $e\alpha _{m} >0$ has the form

\begin{equation} \label{50_} A_{10} \equiv A(M_{10} )={\rm }\left(\begin{array}{cccc} {0} & {1} & {0} & {0} \\ {2e} & {-\sqrt{3\left(\displaystyle\frac{1}{2\alpha _{m} } +\lambda _{m} \right)} } & {0} & {0} \\ {0} & {0} & {0} & {1} \\ {0} & {0} & {\varepsilon \mu ^{2} } & {-\sqrt{3\left(\displaystyle\frac{1}{2\alpha _{m} } +\lambda _{m} \right)} } \end{array}\right), \end{equation}

and its determinant

\begin{equation} \label{51_} \Delta (A_{10} )=2e\varepsilon \mu ^{2} . \end{equation}

The characteristic equation for the matrix $A_{10} $ has the form

\[\left(k^{2} +k\sqrt{3\left(\frac{1}{2\alpha _{m} } +\lambda _{m} \right)} -\varepsilon \mu ^{2} \right)\left(k^{2} +k\sqrt{3\left(\frac{1}{2\alpha _{m} } +\lambda _{m} \right)} -2e\right)=0.\]
and the eigenvalues of the matrix are

\begin{equation} \label{52_} \begin{array}{l} {k_{1} (M_{10} )=-\frac{1}{2} \sqrt{3v} +\frac{1}{2} \sqrt{3v+4\varepsilon \mu ^{2} } };\quad {k_{2} (M_{10} )=-\frac{1}{2} \sqrt{3v} -\frac{1}{2} \sqrt{3v+4\varepsilon \mu ^{2} } ;}\\[12pt]  {k_{3} (M_{10} )=-\frac{1}{2} \sqrt{3v} +\frac{1}{2} \sqrt{3v+8e}} ;\quad \quad {k_{4} (M_{10} )=-\frac{1}{2} \sqrt{3v} -\frac{1}{2} \sqrt{3v+8e} ,} \end{array} \end{equation}

where  ${\rm }v=\displaystyle\frac{1}{2\alpha _{m} } +\lambda _{m} .$

Here

\begin{equation} \label{53_} \begin{array}{l} {k_{1} (M_{10} )\cdot k_{2} (M_{10} )=-\varepsilon \mu ^{2} ;{\rm \; \; }k_{3} (M_{10} )\cdot k_{4} (M_{10} )=-2e;{\rm \; \; }} \\[12pt] {k_{1} (M_{10} )\cdot k_{2} (M_{10} )\cdot k_{3} (M_{10} )\cdot k_{4} (M_{10} )=\Delta (A_{10} ).} \end{array} \end{equation}

Therefore:

\begin{enumerate}
\item  for $e=+1$ $k_{3} $ and $k_{4} $ are real numbers with different signs; for $e=-1$  $k_{3} $ and $k_{4} $ are either complex conjugate numbers or real numbers with identical signs;

\item  for $\varepsilon =+1$ $k_{1} $ and $k_{2} $ are real numbers with different signs; for $\varepsilon =-1$ $k_{1} $ and $k_{2} $ are either complex conjugate numbers or real numbers with identical signs.
\end{enumerate}

\subsection{Characteristic equation and qualitative analysis near the singular points $M_{11} ,M_{12} ,M_{21} ,M_{22} $ }

In order for the dynamical system to allow phase trajectories to arrive at singular points, it is necessary that the coordinates of these points have real values, i.e., that

\[\frac{1}{2\alpha _{m} } +\frac{\mu ^{4} }{2\beta _{m} } +\lambda _{m} \ge 0,\]

The matrix of system \eqref{27}at the singular points \eqref{38} for $e\alpha _{m} >0$, $\varepsilon \beta _{m} >0$  has the form

\begin{equation} \label{54} A_{11} \equiv A(M_{11} )={\rm }\left(\begin{array}{cccc} {0} & {1} & {0} & {0} \\ {2e} & {-\sqrt{3\left(\Frac{1}{2\alpha _{m} } +\Frac{\mu ^{4} }{2\beta _{m} } +\lambda _{m} \right)} } & {0} & {0} \\ {0} & {0} & {0} & {1} \\ {0} & {0} & {-2\varepsilon \mu ^{2} } & {-\sqrt{3\left(\Frac{1}{2\alpha _{m} } +\Frac{\mu ^{4} }{2\beta _{m} } +\lambda _{m} \right)} } \end{array}\right), \end{equation}

and its determinant

\begin{equation} \label{55} \Delta (A_{11} )=-4e\varepsilon \mu ^{2} . \end{equation}

The characteristic equation for the matrix $A_{11} $ has the form

\[\left(k^{2} +k\sqrt{3\left(\frac{1}{2\alpha _{m} } +\frac{\mu ^{4} }{2\beta _{m} } +\lambda _{m} \right)} +2\varepsilon \mu ^{2} \right)\left(k^{2} +k\sqrt{3\left(\frac{1}{2\alpha _{m} } +\frac{\mu ^{4} }{2\beta _{m} } +\lambda _{m} \right)} -2e\right)=0,\]

and the eigenvalues of the matrix are

\begin{equation} \label{56_} \begin{array}{l} {k_{1} (M_{11} )=-\frac{1}{2} \sqrt{3w} +\frac{1}{2} \sqrt{3w-8\varepsilon \mu ^{2} }} ;\quad {k_{2} (M_{11} )=-\frac{1}{2} \sqrt{3w} -\frac{1}{2} \sqrt{3w-8\varepsilon \mu ^{2} } ;} \\[12pt]  {k_{3} (M_{11} )=-\frac{1}{2} \sqrt{3w} +\frac{1}{2} \sqrt{3w+8e} ;\quad \quad k_{4} (M_{11} )=-\frac{1}{2} \sqrt{3w} -\frac{1}{2} \sqrt{3w+8e} ,} \end{array} \end{equation}

where  ${\rm }w=\Frac{1}{2\alpha _{m} } +\Frac{\mu ^{4} }{2\beta _{m} } +\lambda _{m} .$

Here

\begin{equation} \label{57_} \begin{array}{l} {k_{1} (M_{11} )\cdot k_{2} (M_{11} )=2\varepsilon \mu ^{2} ;{\rm \; \; }k_{3} (M_{11} )\cdot k_{4} (M_{11} )=-2e;{\rm \; \; }} \\[12pt]  {k_{1} (M_{11} )\cdot k_{2} (M_{11} )\cdot k_{3} (M_{11} )\cdot k_{4} (M_{11} )=\Delta (A_{11} ).} \end{array} \end{equation}

Therefore:

\begin{enumerate}
\item  for $e=+1$ $k_{3} $ and $k_{4} $ are either complex conjugate numbers or real numbers with identical signs; for $e=-1$  $k_{3}$ and $k_{4}$ are real numbers with different signs;

\item  for $\varepsilon =+1$ $k_{1} $ and  $k_{2} $ are real numbers with different signs; for $\varepsilon =-1$ $k_{1} $ and $k_{2} $ are either complex conjugate numbers or real numbers with identical signs.
\end{enumerate}

\section{Conclusions}

We have confirmed and refined the main conclusions of [1]. At the same time, the main asymptotic properties of the cosmological model based on the asymmetric scalar doublet $\{ \Phi ,\varphi \} $ have become physically more intelligible. In a subsequent paper, we will present and analyze the results of numerical integration of the four-dimensional dynamical system presented above and elucidate the most interesting properties of the model.
The work was performed according to the Russian Government Program of Competitive Growth of Kazan
Federal University.


\end{document}